\documentclass[12pt]{iopart}
\usepackage[utf8]{inputenc}
\usepackage[english]{babel}
\usepackage{graphicx}
\usepackage{subfig}
\usepackage[scale=0.7]{geometry}
\usepackage[figurename=Figure, tablename=Table, labelfont=bf]{caption}
\usepackage[colorlinks]{hyperref}
\usepackage{multicol}
\usepackage{xcolor}
\usepackage{colortbl}
\usepackage{hypcap}
\usepackage{array}
\usepackage{siunitx}
\hypersetup{
 pdfborder={0 0 0},
 breaklinks=true,
 linkcolor=blue,
 menucolor=blue,
 urlcolor=blue,
 citecolor=cyan,
 colorlinks=true,
 pdffitwindow=true,
}
\begin{document}

\title  {A high repetition deterministic single ion source}
\author {C Sahin, P Geppert, A M\"ullers and H Ott}
\address{Department of Physics and Research Center OPTIMAS, Technische Universit\"at 
         Kaiserslautern, Erwin-Schr\"odinger-Strasse, 67663 Kaiserslautern, Germany}
\ead    {\href{mailto:ott@physik.uni-kl.de}{ott@physik.uni-kl.de}}

\begin{abstract}
We report on a deterministic single ion source with high repetition rate and high fidelity. 
The source employs a magneto-optical trap, where ultracold Rubidium atoms are photoionized. 
The electrons herald the creation of a corresponding ion, whose timing information is used to manipulate its trajectory in flight.
We demonstrate an ion rate of up to $4 \times 10^4\,\SI{}{\per\second}$ and achieve a fidelity for single ion operation of 98\,\%.
The technique can be used for all atomic species, which can be laser-cooled, and opens up new applications in ion microscopy, ion implantation and surface spectroscopy.
\end{abstract}

\pacs{29.25.Ni, 68.55.Ln, 85.40.Ry}
\vspace*{10pt}
\noindent{\it Keywords\/}: single-ion source, laser cooling, laser ionization, heralded, deterministic, high repetition\\
\hspace*{-5pt}\submitto{\NJP}
\maketitle

%
%
\section{Introduction}\label{sec:Introduction}

The preparation, manipulation and detection of single particles is one of the major thrusts in physics technology development.
The past decades have evidenced that single particle control \cite{Dehmelt:1980, Rempe:2007, Ott:2016} of ions, atoms or photons is a key enabler for scientific progress. 
The field of applications for ions range from fundamental studies \cite{Vuletic:2009} and quantum optical experiments \cite{Blatt:2013} to technical routines in materials science like nanostructuring \cite{Tseng:2004}, doping \cite{Rogge:2013, Wrachtrup:2009, Lukin:2016}, microscopy \cite{Singer:2016, Poelsema:2014} and surface spectroscopy \cite{Eswara:2015}.
Alongside the established ion sources \cite{Giannuzzi:2014, Gierak:2016}, progress in field ionization sources \cite{Fitzpatrick:2013, Hwang:2017} and new sources based on photoionization of ultracold atoms \cite{McClelland:2016}  extend the range of available species and possible applications.
Only little work has been undertaken so far to control the number of ions in a beam and to trace the space-time trajectory of the individual ions \cite{Scholten:2014}.
Having full control over each single ion in a beam \cite{Singer:2009} would, however, offer a complete new portfolio of ion beam techniques, bridging quantum optical and materials science applications \cite{Jamieson:2015}. 

A variety of ion sources are currently available, each featuring specific properties. Liquid metal ion sources (LMIS), mainly based on Gallium ions \cite{Gierak:2016}, are the workhorse in ion beam applications providing beams with high currents and brightness. More recently, field ionization at a nanotip has enabled a Helium focused ion beam (FIB), which features to date the best spatial resolution of about 0.25\,nm \cite{Wen:2012}.
In these cases, the ions have a broad kinetic energy distributions of a few electronvolts. Due to this large energy spread, ion beam applications typically require beam energies in the 10\,keV range in order to ensure good focusing. Recently, a third type of ion source based on photoionization of laser-cooled atoms is gaining increasing attention \cite{Fuso:2016, Comparat:2017, Wouters:2017}.
Due to the low temperature, such a source can feature a high brightness comparable to a LMIS,  however, at a much lower energy spread of less than 1\,eV, which allows operation also at low beam energies of a few hundred electronvolts \cite{McClelland:2016}. While many experiments so far employed atomic species commonly used in cold gas experiments such as Rubidium, Lithium or Chromium, this method is, in principle, applicable to every element with an electronic transition suitable for laser cooling. \\
In this letter, we extend the concept of a magneto-optical trap ion source (MOTIS) \cite{McClelland:2016} to feature single ion and deterministic operation, where the number of generated ions is known with almost unit fidelity, while at the same time their starting times are measured with very high accuracy. Such a heralded single ion source allows for the delivery of a large number of ions with sub-Poissonian statistics and for the generation of a deterministic string of ions with low time jitter.

%
%
\section{Working principle}\label{sec:WorkingPrinciple}

The source is based on the photoionization of laser-cooled atoms. The created ions are guided through ion optics towards a detector. Heralded operation of this source is achieved by also collecting the electrons from the ionization process and detecting them. Due to their higher velocities, the electrons are detected within tens of nanoseconds after an ionization event and thus provide information about the generation time of the corresponding ions. This information is used to manipulate the ions trajectories in flight with a segmented gate electrode consisting of two half-cylinders, one of which is set to a fixed voltage, while the other half is grounded. By default, the voltage difference is large enough to deflect all ions, corresponding to a closed gate. If an electron is detected, a voltage pulse to the grounded half-cylinder equals the voltages on both segments, allowing the corresponding ion to pass. The pulse must be applied after a delay time matching the ions time of flight to the gate electrode. Ions that reach the gate electrode before or after the switching pulse are thus deflected. This simple but powerful concept ensures that every ion in the beam is heralded by the detected electron. Figure \ref{fig:ExpermentalSetup} illustrates this basic working principle.

%
%
\section{Experiment}\label{sec:Experiment}
\begin{figure}[ht!]
  \centering
  \includegraphics[scale=1]{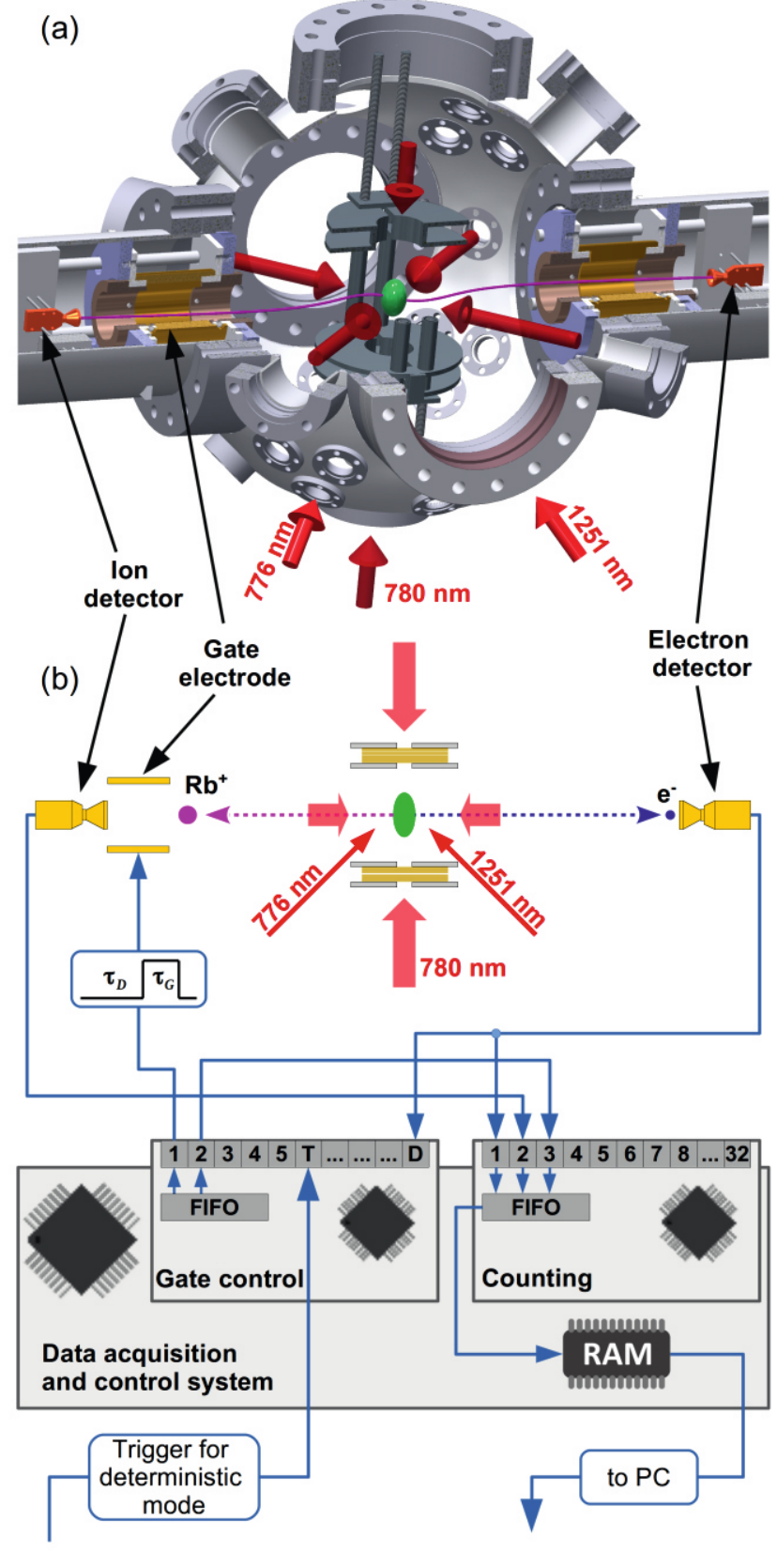}
  \caption{Setup. (a) Experiment chamber featuring a rubidium MOT  and two symmetric detector arms, each with a CEM placed approximately $\SI{22}{cm}$ from the center. The MOT is loaded from background vapor. The $\SI{776}{nm}$ and $\SI{1251}{nm}$ laser beams form an ionization volume of $\SI{200}{\micro\metre}$ diameter, in which laser-cooled atoms in the excited state are photoionized in a two-photon process. Electrons and ions are extracted in opposite directions and detected with two CEMs. (b) Scheme of the experiment control with two real-time processors used for particle counting and gate control. For details, see appendix. }
  \label{fig:ExpermentalSetup}
\end{figure}

In our experimental setup (see fig. \ref{fig:ExpermentalSetup}) we use an ultracold cloud of $^{87}$Rb atoms. Beside the MOT \cite{Pritchard:1987} lasers (cooling and repumping light at $\SI{780}{nm}$), another two infrared lasers ($\SI{776}{nm}$ and $\SI{1251}{nm}$) are used to create electron-ion pairs in a three-photon ionization process. We use infrared lasers only to minimize the rate of photoelectrons, which, when reaching the electron detector, would trigger gate pulses without corresponding ions.
Both the $\SI{776}{nm}$ and $\SI{1251}{nm}$ lasers are focused down to about $\SI{200}{\micro\metre}$ and overlap perpendicularly inside the MOT volume. The ionization fragments are extracted with electric fields and guided towards two CEM detectors positioned at a distance of about $\SI{22}{cm}$ on opposite sides of the ionization volume. The electron travels this distance within $\SI{60}{ns}$, while the ion's time of flight is -- depending on the applied electric fields -- three orders of magnitude larger ($\SI{30}{} - \SI{50}{\micro\second}$). The recorded electron heralds the created ion, which can then be individually accepted or deflected with the segmented electrode (gate electrode). Appropriate settings for the operation of the gate electrode are found through numerical simulations and verified experimentally.

We can classify the ion source into continuous, single ion and deterministic operation modes, depending on the setting of the gate control. When the gate control is disabled, allowing all ions to pass, we record ion rates in continuous mode of up to $1\times10^{6}\;\SI{}{\per\second}$, which is the limit of our data acquisition system. Single ion operation is enabled by triggering the gate control conditional to a detected electron.
Deterministic operation is realized by an additional trigger channel to the gate control, which determines when the next ion should be accepted. 

The experiment is controlled by a dedicated real-time system with two separate processors for event counting and gate control (see fig. \ref{fig:ExpermentalSetup}(b) and appendix). For deterministic operation, the requested pattern is generated by a function generator in burst mode.

%
%
\section{Results}\label{sec:Results}

We characterize the source in terms of two key properties. In single ion and deterministic mode, the source should, ideally, only emit a single ion at a time and each ion should be attributed to a single gate switching cycle. This is expressed in terms of the fidelity. Additionally, in deterministic mode, the time jitter, describing the variation in arrival times after an ion request, is of importance.

\begin{figure}[ht!]
  \centering
  \includegraphics[scale=1]{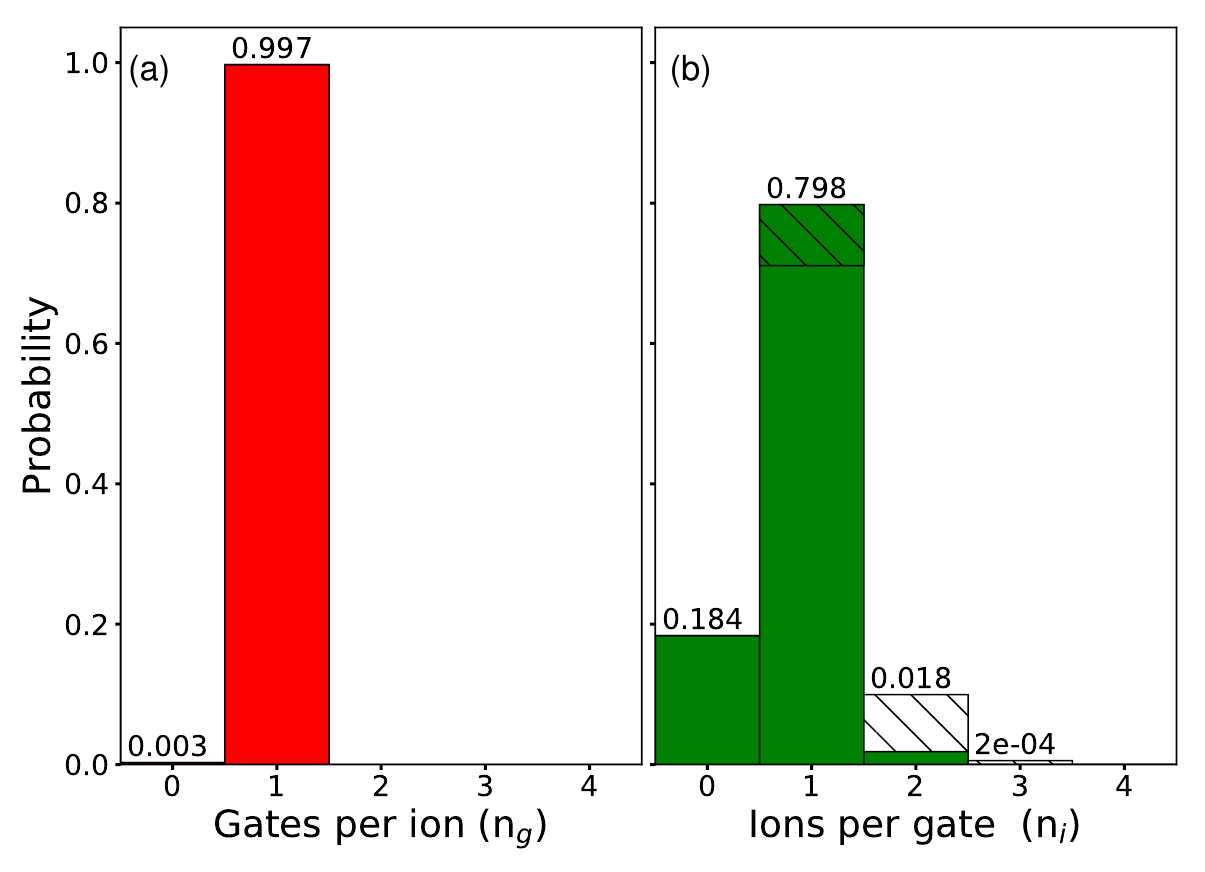}
  \caption{ Probability distributions for correlated ion-gate pairs. Sample dataset for an ion rate of $\SI{11\,500}{\per\second}$. (a) Probabilities for detecting $n_g$ gates in coincidence with a single ion $(n_i=1)$. $\SI{99.7}{\percent}$ of ions can be attributed to a single gate. The remaining $\SI{0.3}{\percent}$ of ions with no corresponding gate are caused by dark counts in the ion detector and stray ions.  (b) Probabilities for detecting $n_i$ ions in coincidence with a single gate $(n_g=1)$. The $n_i = 0$ bin sums up ions missed due to limited detection efficiency as well as gates triggered by background electrons (see text). Entries with $n_i \geq 2$ correspond to multiple ions passing through a single gate. The hatching indicates the distribution before filtering is applied, for details see text and appendix.}
  \label{fig:ConditionalProbability}
\end{figure}

We extract the fidelity from a correlation analysis of the recorded ion and gate signal in a properly chosen coincidence window (see fig.\,\ref{fig:ConditionalProbability} and appendix). The fidelity can then be written as: 

\begin{equation}
\hspace*{-50pt}
\text{F} = 1 - \text{P}(n_i>1|n_g=1) - \text{P}(n_g>1|n_i=1) - \text{P}(n_g=0|n_i=1) - \text{P}_{bg,g} 
\end{equation}
Here, $\text{P}(n_i>1|n_g=1)$ is the conditional probability for finding more than one ion ($n_i>1$) in coincidence with a single gate $(n_g=1)$. Accordingly, $\text{P}(n_g>1|n_i=1)$ and $\text{P}(n_g=0|n_i=1)$ correspond to the cases where a single ion $(n_i=1)$ may be attributed to multiple gates or no matching gate can be found, respectively. 
In the correlation analysis, we always find a fraction of about 18\,\% of all events where a gate has no corresponding ion, $\text{P}(n_i=0|n_g=1)$. This is due to the finite efficiency of the ion detector and does not contribute to the fidelity, apart from a small contribution, wich stems from background electrons. 
This contribution is denoted as $\text{P}_{bg,g}$ and is measured separately without the ionization lasers.
We find a typical background electron rate of $80\pm40\,\SI{}{\per\second}$. 
For high ion rates, background electrons can be neglected. 
For low ion rates however, they are the dominating limiting factor of the source performance.

When analyzing the time intervals between consecutive events, we find that for both electrons and ions the fraction of closely spaced events is considerably larger than expected from a Poissonian process like laser ionization. Since the effect is also present when recording background photoelectrons only, we attribute this to a technical cause, most likely detector ringing. Since they cause a significant underestimation of the fidelity, we apply a post-run filtering of our
data to remove these unphysical events (see appendix). The histogram in figure \ref{fig:ConditionalProbability}(b) shows the distribution after filtering, the corresponding values before filtering are indicated by the hatching.

\begin{figure}[ht!]
  \centering
  \includegraphics[scale=1]{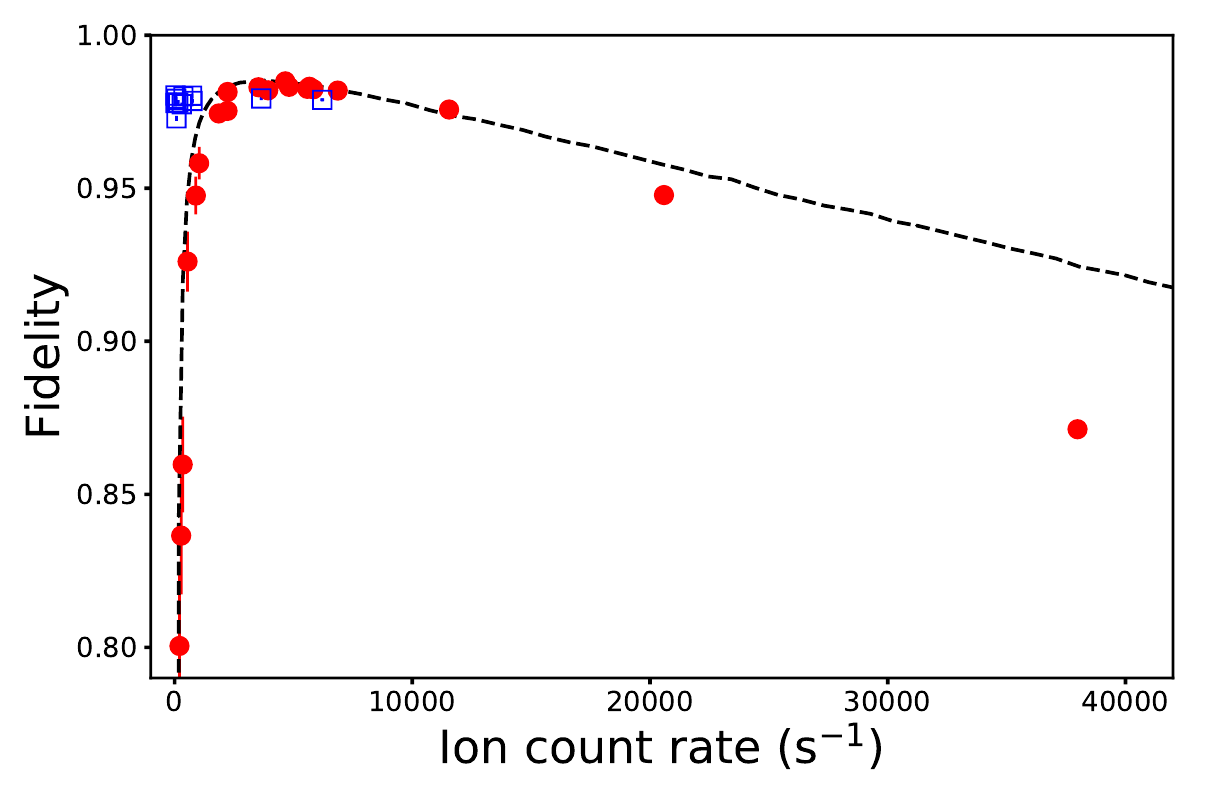}
  \caption{Fidelity of the source in dependence of the ion rate. The red circles correspond to single ion operation mode. The dashed black line shows the expected fidelity (see text). Blue squares depict the fidelity in deterministic operation mode.}
  \label{fig:Fidelity_vs_Ionrate}
\end{figure}

\clearpage
The fidelity for different ion rates is shown in figure \ref{fig:Fidelity_vs_Ionrate}.
We observe a maximum fidelity of about $\SI{98}{\percent}$ within a plateau ranging from 500 to $\SI{5000}{\per\second}$. 
For lower count rates, the constant electron background rate causes a drop in the fidelity, while the decrease at larger ion rates is caused by the increasing probability of multiple ions passing through a single gate. The dashed black line shows the highest achievable fidelity for an electron background rate of $80\,\SI{}{\per\second}$ and a probability for two ions per gate governed purely by a Poisson distribution in the ion statistics. For almost all count rates, our results closely match the expected curve, showing that these two contributions are the dominant factors. Only for higher count rates, the measured fidelity drops below the curve. The reason for this may be three or more ions passing through a gate or an incomplete filtering of detector ringing.
Our single ion source features strong subpoissonian statistics: A fidelity of 98\,\% means that the source can deliver up to 2500 ions, before the standard deviation of the ion distribution equals that of a stochastic source.

\begin{figure}[ht!]
  \centering
  \includegraphics[scale=1]{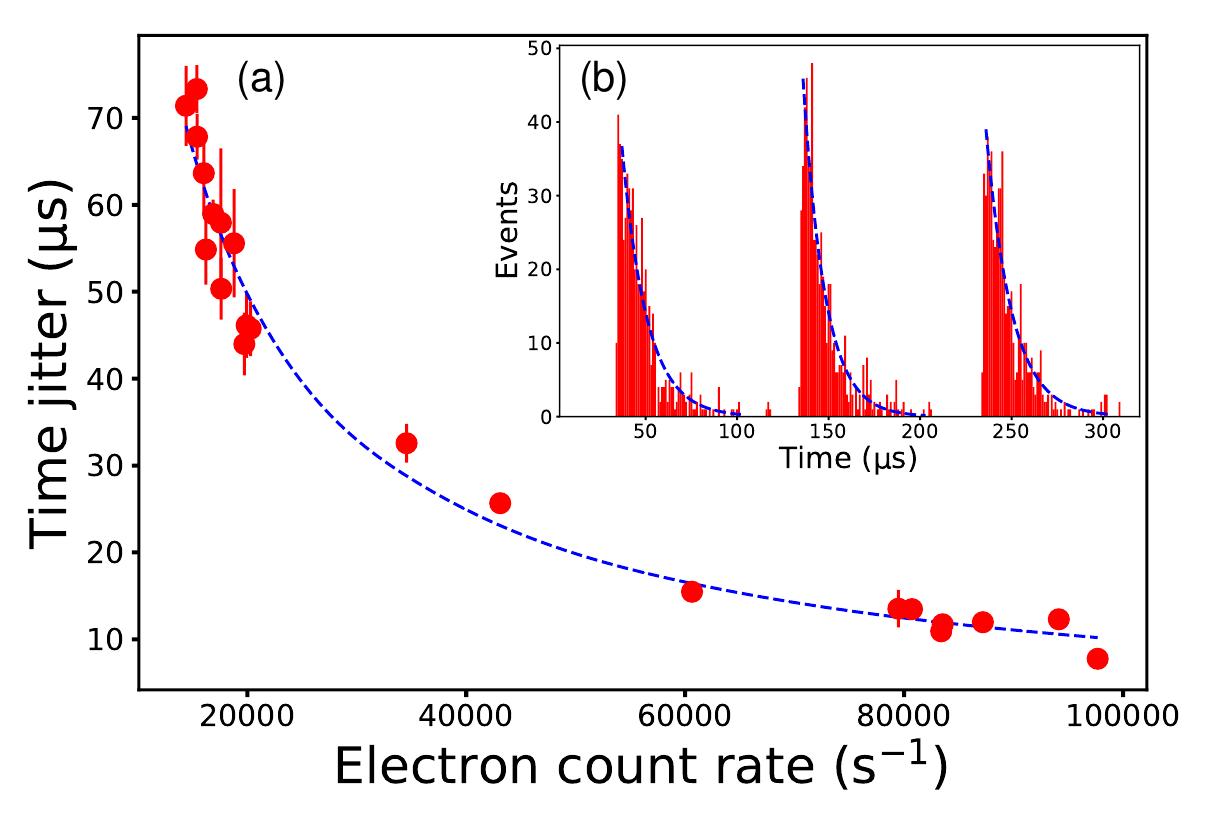}
  \caption{Deterministic operation mode. (a) The plot shows the decrease in time jitter with growing electron count rate. The dashed blue line is a fit with an $1/\text{rate}$ dependence. (b) Distribution of deterministic ion time stamps for a requested ion rate of $\SI{10000}{\per\second}$. We use the 1/e-time from exponential decay fits (dashed blue lines) to quantify the jitter.
  }
  \label{fig:TimeJitter_vs_Electronrate}
\end{figure}

In deterministic mode, after receiving a request, the source emits the next available ion. The time distribution for three consecutive ions is depicted in figure  \ref{fig:TimeJitter_vs_Electronrate}(b) for a requested deterministic rate of $\SI{10000}{\per\second}$ with equal spacing (the underlying electron count rate of the source was $\SI{80000}{\per\second}$).
Compared to previous experiments with single ions emitted from a Paul trap \cite{Singer:2016,Singer:2009}, this is an increase of more than three orders of magnitude. 
The stochastic nature of the ionization process leads to an exponential decay of the emission time.
The emission pattern itself is exactly repeated every $\SI{100}{\micro\second}$.
The time constant of the decay quantifies the variation in emission times, which we call the jitter.
Figure \ref{fig:TimeJitter_vs_Electronrate}(a) shows that the jitter decreases inversely proportional to the ionization rate. For the largest rates achievable in our experiment the jitter is as small as $\SI{10}{\micro\second}$.
Together with a fidelity of almost unity, our source offers superb conditions for applications with deterministic ion production.

%
%
\section{Outlook}\label{sec:Outlok}

We close this work with an outlook on potential applications of a heralded or deterministic single ion source. First we note that the operation principle can be extended to any atomic species, which can be laser-cooled. In fact, it may even be applicable to gas targets at room temperature, provided a suitable multi-photon ionization scheme  and reliable electron detection is available. 
In deterministic operation, a controlled number of particles can be doped into a host matrix \cite{McCallum:2016}. In case of erbium, this allows a controlled doping of miniaturized regions for waveguide design \cite{Kenyon:2005} or the creation of color centers \cite{Rogge:2013}.
Heralded operation in microscopy gives a handle to increase the signal to noise ratio with help of a gated secondary electron detector \cite{Singer:2016}. For microscopy applications, lithium is a well suited candidate \cite{McClelland:2014}. 
In surface spectroscopy, having access to the impact time for each individual projectile ion could allow for the measurement of the energy loss of the scattered ions with high precision in a time-of-flight setting \cite{Winograd:2011}. This can tremendously increase the sensitivity.
Also, the properties of such a source could be beneficial for the demonstration of single ion interferometry \cite{Hasselbach:2010, Stibor:2015}.
Another application could be the correction for image aberrations in charged particle optics.
When the starting time and position of an ion in a beam is known, its trajectory is fully determined by the electric fields of the ion optics. Usually, lenses for charged particles cannot defocus the beam apart from non-rotationally symmetric field configurations. Using dynamical lens fields, where the focal length of the lens is changed while the ion is passing, allows to invert the sign of the spherical aberration coefficient, thus correct imaging aberrations. 
This list is far from being complete and we expect heralded and deterministic ion sources to become an integral part of future ion technology.

%
%
\clearpage
\ack
The authors thank Carsten Pfleger and Torsten Manthey for constructing the MOT apparatus and Jens Benary for building the $\SI{776}{nm}$ and $\SI{1251}{nm}$ lasers. We gratefully acknowledge financial support by the German Science Foundation (DFG) within the OT 222/3-1.

%
%
\section*{Appendix}\label{sec:Appendix}

\subsection*{Laser ionization}\label{sec:Laserionization}
We employ a three photon ionization scheme, which can be implemented using only diode lasers in the infrared spectrum and therefore greatly reduces the rate of photoelectrons from surfaces.
For the first excitation step, we make use of the cooling transition required for the MOT. $^{87}$Rb atoms are excited from the $5\text{S}_{1/2}$ ground state to the $5\text{P}_{3/2}$ excited state by a home built distributed feedback (DFB) diode laser emitting light with a wavelength of $\SI{780}{nm}$. The cooling light is red detuned by $\SI{16.9}{MHz}$ from the $F=2 \rightarrow F^{\prime}=3$ hyperfine transition. The laser light is amplified by a tapered amplifier (TA) with a maximum output power of $\SI{2}{W}$. The repumping transition $F=1 \rightarrow F^{\prime}=2$ for the MOT is driven by a second identical DFB laser.

From the $5\text{P}_{3/2}$ state, the Rubidium atoms are further excited to the $5\text{D}_{5/2}$ state. The required laser light with a wavelength of $\SI{776}{nm}$ is provided by an external cavity diode laser (ECDL) in Littrow configuration. Its frequency is locked by Doppler free saturation spectroscopy in a heated vapor cell, where the pump beam is provided by the MOT cooling laser. Before entering the experiment, the laser beam is sent through an acousto-optical modulator (AOM), which can be used to switch the ionization process from our experiment control system. Less than $\SI{1}{mW}$ of optical power is required to drive the transition.

Ionizing the atoms from the $5\text{D}_{5/2}$ state requires photons with a wavelength of $\SI{1251}{nm}$, which are provided by a second ECDL. We tune the laser frequency to a value just above the ionization threshold, which allows us to omit frequency stabilization. The ionization rate is adjusted by tuning the optical power of this laser from a few hundred microwatts to $\SI{10}{mW}$. While the cooling beams are collimated to a diameter of $\SI{12}{mm}$, the $\SI{776}{nm}$ and $\SI{1251}{nm}$ laser beams are focused into the chamber creating an ionization region of about $\SI{200}{\micro\metre}$ diameter.

\subsection*{Experiment control}\label{sec:ExperimentControl}
The experiment is controlled by a commercial real-time operating system with exchangeable modules for analog and digital in- and output (see fig. \ref{fig:ExpermentalSetup}(b)). The main processor controls the experimental sequences and transfers acquired data to a PC. Two of the extension modules feature separate real-time processors, which in turn can execute independent tasks. The first processor, running with a clock frequency of $\SI{100}{MHz}$, collects events from both detectors using a first-in-first out (FIFO) hardware buffer. The FIFO is periodically read out by the processor, which stores the events in the device RAM from where the data is subsequently transferred to a PC.

The second processor runs with a higher clock frequency of $\SI{200}{MHz}$ and is used for gate control. The signal from the electron detector is split and triggers a readout process, which stores the event time in local memory.
The process discards events which would lead to overlapping gate cycles.
The gate opening and closing times are programmed into a hardware buffer (FIFO in output configuration), from where they are executed once the module clock reaches the designated time. A TTL level voltage pulse is sent to a remotely controllable power supply, which in turn is connected to the segmented gate electrode. In addition, a second TTL pulse of $\SI{30}{ns}$ duration is sent to the first processor where it is counted as a third detector channel for monitoring purposes.

In deterministic mode, the readout process collects events as above, however, no gates are programmed by default. A second process listens for ion requests on a trigger channel. If a signal is received, an internal state variable is changed, indicating to the readout process to let the next available ion pass. After programming the corresponding gate, the readout process resets the state variable.

\subsection*{Filtering}\label{sec:Filtering}
Throughout our experiments, we observe that the probability for multiple ions in coincidence with a single gate is considerably larger than the value expected from a Poissonian process. This excess of closely timed events is present in both electron and ion channels and shows up when running the source both with and without gate operation and also when recording background photoelectrons only.
The effect strongly depends on the voltages applied to the CEM detectors and the electrodes and can therefore be attributed to a technical cause, most likely detector ringing.
To remove these unphysical events, we take for each experiment a corresponding reference dataset without gate operation. We then bin the time differences between events and randomly remove entries until the dataset matches the expected distribution. The same fraction of events per time bin is then removed from the corresponding dataset with gate operation. On average, $\SI{5}{\percent}$ to $\SI{10}{\percent}$ of the events are filtered. 

\clearpage
\section*{References}
\bibliographystyle{unsrt}
\bibliography{Literature}

\end{document}